\documentclass[prl,twocolumn,a4paper,groupedaddress,showpacs,floatfix]{revtex4}
\usepackage{graphicx,amsmath,amssymb,amsfonts,dsfont,subfigure,tikz,hyperref}

\usepackage{color}

\newcommand{\Title}{Asymmetry in energy versus spin transport in certain interacting, disordered systems}
\begin{document}

\title{\Title}

\author{J. J. Mendoza-Arenas$^{1,2,*}$, M. Znidaric$^{3}$, V. K. Varma$^{4,5}$, J. Goold$^{6}$, S. R. Clark$^{7,8}$, A. Scardicchio$^{9,10}$}

\affiliation{${}^1$Departamento de F\'{i}sica, Universidad de los Andes, A.A. 4976, Bogot\'a D. C., Colombia}
\email{ jj.mendoza@uniandes.edu.co}
\affiliation{${}^2$Clarendon Laboratory, University of Oxford, Parks Road, Oxford OX1 3PU, United Kingdom}
\affiliation{${}^3$Physics Department, Faculty of Mathematics and Physics, University of Ljubljana, Ljubljana, Slovenia}
\affiliation{${}^4$College of Staten Island and Graduate Center, CUNY, New York, USA}
\affiliation{${}^5$Department of Physics and Astronomy, University of Pittsburgh, PA, USA}
\affiliation{${}^6$School of Physics, Trinity College Dublin, Dublin 2, Ireland}
\affiliation{${}^7$Department of Physics, University of Bath, Claverton Down, Bath BA2 7AY, UK}
\affiliation{${}^8$Max Planck Institute for the Structure and Dynamics of Matter, University of Hamburg CFEL, Germany}
\affiliation{${}^9$Abdus Salam ICTP, Strada Costiera 11, 34151 Trieste, Italy}
\affiliation{${}^{10}$INFN, Sezione di Trieste, Via Valerio 2, 34126 Trieste, Italy}


\date{\today}

\begin{abstract}
We study energy transport in $XXZ$ spin chains driven to nonequilibrium configurations by thermal reservoirs of different temperatures at the boundaries. 
We discuss the transition between diffusive and subdiffusive transport regimes in sectors of zero and finite magnetization at high temperature. 
At large anisotropies we find that diffusive energy transport prevails over a large range of disorder strengths, 
which is in contrast to spin transport that is subdiffusive in the same regime for weak disorder strengths.
However, when finite magnetization is induced, both energy and spin currents decay as a function of system size with the same exponent.
Based on this, we conclude that diffusion of energy is much more pervasive than that of magnetization in these disordered spin-1/2 systems, and occurs across a significant range of the 
interaction-disorder parameter phase-space; we suggest this is due to conservation laws present in the clean $XXZ$ limit. 
\end{abstract}

\maketitle

\textit{\underline{Introduction}}: What determines transport of conserved quantities in generic one-dimensional disordered systems? 
In contrast to classical systems, where diffusion is prevalent, there is no universal answer to this question for quantum systems \cite{Anderson:1958} where nonstandard hydrodynamical behaviours emerge frequently \cite{ChaikinLubensky, Forster}.
\begin{figure}[tph!]
\centering
\includegraphics[width=0.4\textwidth]{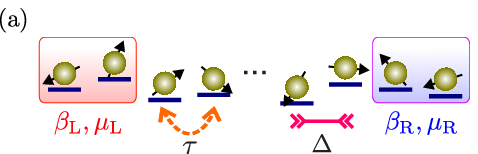}
\includegraphics[scale=0.6]{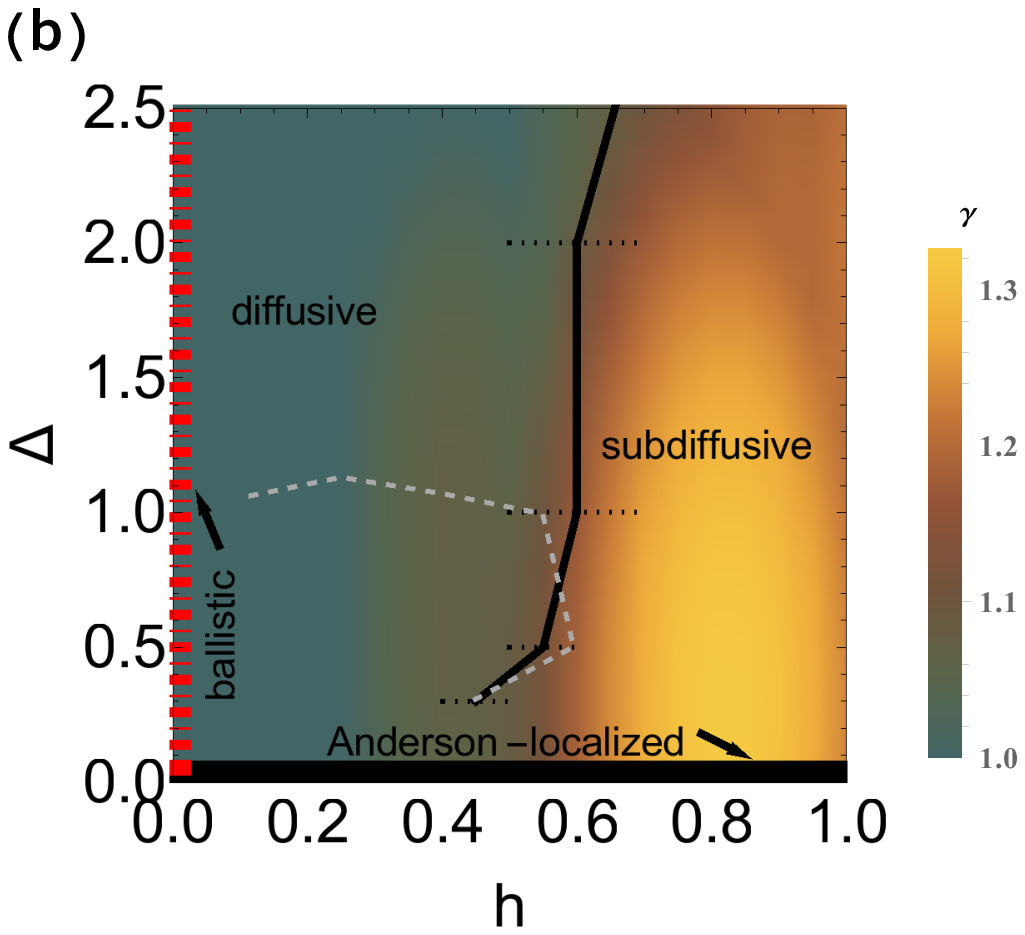}
\caption{(a) Scheme of a disordered $XXZ$ spin-$1/2$ chain driven out of equilibrium by unequal boundary reservoirs, which impose a temperature and/or chemical potential imbalance by inducing different 
two-site thermal states at each edge.
(b) Energy transport dynamical phase diagram, indicated by the current scaling exponent $\gamma$ as a function of interaction $\Delta$ and disorder $h$. 
The solid black line defines the boundary between diffusive (to the left) and subdiffusive (to the right) conduction in the zero magnetization sector, and the dotted horizontal lines correspond to the 
error bars of the boundary. 
The underlying colours are Gaussian extrapolation of data, with brighter colours ($\gamma > 1$, orange) indicating subdiffusive transport and darker colours ($\gamma=1$, green) indicating normal transport.
The gray dashed line is the corresponding diffusive-subdiffusive crossover for spin transport in the same sector (see Ref. \cite{Znidaric:2016}). 
%We also indicate the ballistic (thick dashed-dotted red vertical line at $h=0$) and the Anderson localized (thick solid black horizontal line at $\Delta=0$) limits.
}
\label{enerdiagram_fig}
\label{setup_fig}
\end{figure}
In the present paper, we address this by focussing on a well-known model of interacting spins (so-called disordered $XXZ$ model) where the effect of weak disorder has been shown to result in 
a slow propagation of excitations \cite{Agarwal2015, Lev:2015, Luitz:2016, Varma:2017JSTAT, Znidaric:2016, Marko:2017ann}. 
Moreover, for larger values of disorder, all forms of transport in this model vanish, and the system transits into a \emph{many-body localized} phase, 
where ergodicity is lost in favour of a robust integrable phase \cite{Steinigeweg:2006, Basko:2006, Oganesyan:2007, Pal:2010, abanin2017recent, agarwal2017rare, imbrie2017local}. 

The slow dynamics can occur in both spin \cite{Lev:2015, Agarwal2015, Luitz:2016, Znidaric:2016, Marko:2017ann} 
as well as energy \cite{Varma:2017JSTAT, vosk:2015,potter:2015,Gopalakrishnan:2016} transport (the only two conserved quantities of the model). 
Although the transport of different quantities has been suggested before to be different \cite{Varma:2017JSTAT, Gopalakrishnan:2016}, this has not been unambiguously demonstrated. 
In order to settle this question, we employ a novel technique of coupling spin baths at the ends of an archetypal interacting-disordered system to drive energy or spin currents or both; 
with this approach we are able to access large system sizes $L \approx \mathcal{O}(10^2)$ and thereby clearly unveil \textit{the pivotal role of interactions} 
in establishing the asymmetry between energy and spin transport in the system.

Anomalous transport (either faster or slower than diffusive) in quantum mechanical Hamiltonian systems has been associated with various other properties: 
eigenfunction and spectral fractality \cite{Piechon:1996, Geisel:1997, Varma:2017}, presence of conservation laws or their approximate emergence due to frustrating dynamical constraints 
\cite{zotos:1997,prosen:2011,prosen:2013}, and has been ascribed to the presence of rare regions \cite{agarwal2017rare} in disordered systems. 
In the clean $XXZ$ spin-1/2 chain, which has an infinite set of conserved operators, the energy current operator is a conserved quantity whereas the spin current operator is not \cite{zotos:1997}. This implies that the energy current does not decay, with its current-current correlation function persisting to a plateau in the long time 
limit, giving rise to a Drude peak characteristic of ballistic transport %Therefore ballistic transport of energy is present for any value of interaction strength in the chain, for both zero and nonzero total magnetization.
\cite{zotos:1997,klumper:2002,orignac:2003,louis:2003,Heidrich-Meisner:2003,heidrich:2007,langer:2011,karrasch:2013,karrasch:2015,Mendoza:2015}.

The situation for spin conduction in the clean $XXZ$ is altogether different: while it is not a conserved quantity, 
Mazur's inequalities \cite{Mazur_ineq, zotos:1997} together with the conserved quantities in the model may be invoked to show that high-temperature transport is indeed ballistic for anisotropy $\Delta < 1$ \cite{prosen:2011, prosen:2013}. On the other hand, evidence of superdiffusive transport for $\Delta = 1$ and of diffusion for $\Delta > 1$ has been obtained from numerical simulations \cite{McCoy1997,Benenti:2009,Znidaric:2011b,Mendoza:2013a,Mendoza:2013b,Znidaric:2016,Ljubotina,OganesyanVarma:2017,SanchezVarma:2017}.
In the nonzero magnetization sector however spin transport is always ballistic due to finite overlap of the spin currents with conserved quantities. 

In the present work we find that these clean system properties strongly influence the energy and spin transport dynamics upon the introduction of disorder. 
In particular, our results demonstrate that if energy and spin are transported differently (or similarly) in the clean limit, 
then this relationship is retained in the disordered chain, and must therefore be considered as a true, physical characteristic of the system's dynamics. Surprisingly this holds even for disorder strengths $h/J = \mathcal{O}(1)$, with $J$ setting the energy scale of the Hamiltonian, where we would expect clean system physics to be washed out by the disorder, together with any fingerprints of its integrability. 
%Rather, the transport of different quantities of the clean system, which is usually associated with integrability, survives the breakdown of this integrability and must therefore be considered as a true, physical characteristic of the system's dynamics.

\textit{\underline{Model and Method}}: Here we describe the nonequilibrium setup used to study the energy transport across a disordered quantum system, depicted in Fig.~\ref{setup_fig}(a). We consider a 1D spin-$1/2$ lattice modeled by a $XXZ$ Hamiltonian with a disordered magnetic field along the $z-$ axis
\begin{align} \label{hamiltonian}
\begin{split}
H=\sum_{i=1}^{L-1}&\Bigl[\tau\bigl(s_i^xs_{i+1}^x+s_i^ys_{i+1}^y+\Delta s_i^zs_{i+1}^z\bigr)\\
&+\frac{h_i}{2}s_i^z+\frac{h_{i+1}}{2}s_{i+1}^z\Bigr]=\sum_{i=1}^{L-1}\varepsilon_{i,i+1}.
\end{split}
\end{align} 
Here $L$ is the number of sites in the chain, $s_i^{\alpha}=\frac{1}{2}\sigma_i^{\alpha}$ ($\alpha=x,y,z$) are spin-$1/2$ operators for site $i$ (and $\sigma_i^{\alpha}$ the Pauli matrices, taking $\hbar=1$ throughout the manuscript), $\tau$ is the exchange interaction between nearest neighbors, $\Delta$ is the anisotropy along $z$ direction, $h_i\in[-h,h]$ is the uniformly-random magnetic field at site $i$, and $h$ is the strength of the disorder. 
%A Jordan-Wigner transformation links this Hamiltonian to a model of spinless fermions, where $\tau$ corresponds to the hopping between nearest neighbors and $\tau\Delta$ to a density-density interaction. 
In the following we set the energy scale by taking $\tau=1$.

To induce a temperature and/or chemical potential imbalance across the lattice, we assume that both its left ($k=\text{L}$) and right ($k=\text{R}$) boundaries are coupled to a reservoir characterized by an inverse temperature $\beta_k$ and a chemical potential $\mu_k$, 
as depicted in Fig.~\ref{setup_fig}(a). The dynamics of its density matrix $\rho$ is governed by the Lindblad master equation
\begin{equation} \label{lindblad}
\mathcal{L}\rho=\frac{d\rho}{dt}=-{\rm i}[H,\rho]+\mathcal{L}_{\text{L}}(\rho)+\mathcal{L}_{\text{R}}(\rho).
\end{equation}
$\mathcal{L}$ is the total propagator of the system; the commutator corresponds to the coherent dynamics, and the terms $\mathcal{L}_{k}$ represent the incoherent action of the reservoirs on the chain. 
Reservoir $k=\text{L}$(R) is directly coupled to the two left-most (right-most) spins of the chain, in such a way that if they were separated from the rest, a two-site NESS $\rho_{\text{L}}$($\rho_{\text{R}}$), corresponding to a reduced Gibbs state at an inverse temperature $\beta_k$, would be induced on them. These \textit{target} states $\rho_k$ are defined in the Supplemental Material~\cite{Suppl}.

To characterize the nature of the spin and energy transport we discuss the spin and energy profiles, and the steady state local current values. These correspond to the expectation values of the local magnetization $\langle s_i^z\rangle$ and energy density $\langle\varepsilon_{i,i+1}\rangle$ for all values of $i$, and of the local current operators obtained from continuity equations~\cite{zotos:1997}. 
The magnetization current is $j_i^{\text{S}}=(s_i^xs_{i+1}^y-s_i^ys_{i+1}^x)$, and the energy current is
$j_i^{\rm E}=[(s_{i-1}^ys_i^zs_{i+1}^x-s_{i-1}^xs_i^zs_{i+1}^y) + \Delta(s_{i-1}^zs_i^xs_{i+1}^y-s_{i-1}^ys_i^xs_{i+1}^z) + \Delta(s_{i-1}^xs_i^ys_{i+1}^z-s_{i-1}^zs_i^ys_{i+1}^x)+(h_i/2)(j_i^{\text{S}}+j_{i+1}^{\text{S}})]$.

These observables are determined for several disorder realizations, over which the average is performed. 
The NESS for each realization, defined as $\rho_{\infty}=\lim_{t\rightarrow\infty}\exp(\mathcal{L}t)\rho(0)$, 
is obtained with the time-dependent density matrix renormalization group (t-DMRG) algorithm~\cite{Zwolak:2004,Cirac:2004,Prosen:2009,Schollwock:2011,tnt_review1,tnt}.
In the NESS both currents are homogeneous, so their disorder-averaged values are simply denoted by $j^{\alpha}$ $(\alpha=\text{S,E})$.
Details on the simulations and obtaining the NESS are contained in the Supplemental Material~\cite{Suppl}.

The diffusion equation for transported quantity $A_{\alpha}$, with $A_{\text{S}}=\langle s_i^z\rangle$ and $A_{\text{E}}=\langle\varepsilon_{i,i+1}\rangle$, is given by 
$j^{\alpha} = -D_{\alpha}\nabla A_{\alpha} =-D_{\alpha}\Delta A_{\alpha}/L$, with $D_{\alpha}$ being the corresponding diffusion constant, $\nabla A_{\alpha}$ the gradient of $A_{\alpha}$ across the chain, and  $\Delta A_{\alpha}$ the difference between its boundary values. 
The general NESS current scaling gives
$j^{\alpha}\sim\frac{1}{L^{\gamma}}$.
Here $\gamma = 1$ corresponds to normal diffusive transport i.e. Fick's law. 
When Fick's law breaks down, there is no longer diffusion in the system and the transport may be slower (subdiffusive, $\gamma > 1$) or quicker (superdiffusive, $\gamma < 1$). 
%
%For ballistic transport ($\gamma = 0$) there is a nondissipative channel surviving in the thermodynamic limit, resulting in a nondecaying finite current independent of system size.
%
The scaling of the NESS current will be our primary diagonostic for inferring dynamics in this driven many-body system. We first study the case of zero magnetization, for which the main result corresponding to the diffusive-subdiffusive energy transport phase diagram is indicated in Fig.~\ref{setup_fig}(b). Then we discuss the scenario of nonzero magnetization.

\begin{figure}[tp!]
\centering
\includegraphics[width=0.42\textwidth]{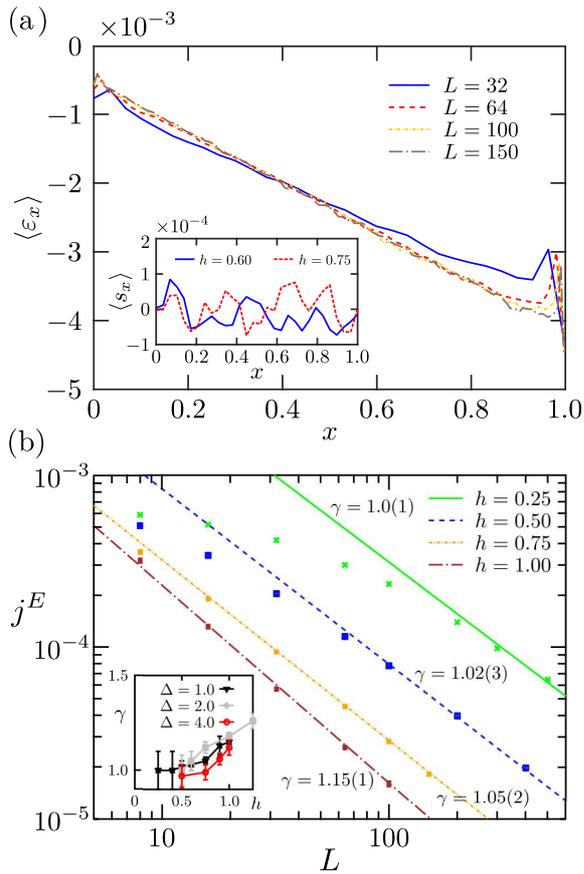}
\caption{NESS transport properties at zero magnetization for $\Delta=1$ and various disorder strengths $h$. (a) Spatial energy profiles across the spin chain for different system sizes, $h=0.60$ and $\Delta=1$, with scaled axis $x=(k-1)/(L-2)$ for $k=1,\ldots,L-1$. Inset: Magnetization profiles for $L=32$ and different disorder strenghts, with scaled axis $x=(k-1)/(L-1)$ for lattice sites $k=1,\ldots,L$. (b) Scaling of energy current for $\Delta=1.0$; the disorder strength $h$ increases from top to bottom. The symbols correspond to the results of the simulations, and the lines are fits to the scaling $j^{\text{E}}\sim L^{-\gamma}$. 
%The obtained values of $\gamma$ are indicated next to each curve. 
Inset: Scaling exponents $\gamma$ as a function of $h$ for different anisotropies $\Delta$. 
%At small $h$ values the data indicate a linear scaling $j \sim L^{-1}$ i.e. diffusion, which becomes subdiffusive (steeper scaling of current $j$ with system size $L$) upon increasing $h$.
}
\label{enerscalings}
\end{figure}

\textit{\underline{Transport at zero magnetization}}: We first consider the NESS transport properties of the spin chain when the boundary driving imposes a finite energy current and a negligible spin current (see Ref.~\cite{Suppl} for details). This corresponds to an energy gradient across the lattice and almost zero total magnetization with no bias between the boundaries, as shown in the main panel and the inset of Fig.~\ref{enerscalings}(a) respectively.    

\begin{figure}[t]
\centering
\includegraphics[width=0.38\textwidth]{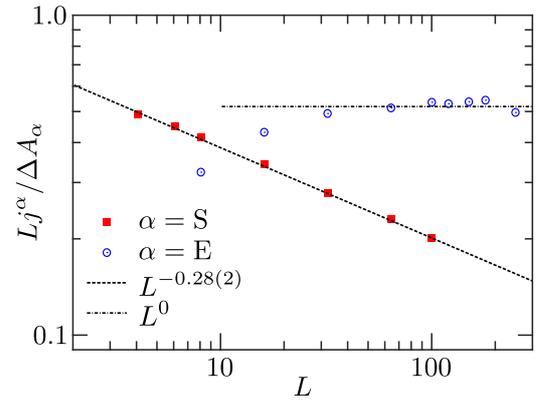}
\caption{NESS transport at zero magnetization, showing differences in energy and spin transport in the strongly-interacting case $\Delta=1.5$ and $h=0.6$ showing diffusive and subdiffusive transport respectively. 
This is in qualitative agreement with the conclusions of Ref. \cite{Varma:2017JSTAT} that spin and energy transport can be different in this model.}
\label{Seebeck_currents_D175}
\end{figure}

In the main panel of Fig. \ref{enerscalings}(b) we show the scaling of NESS current for $\Delta=1$ and a range of disorder amplitudes; we see that diffusion persists up to a finite disorder strength, much like in the 
case of spin transport \cite{Znidaric:2016}. Then at some critical field $h_c \approx 0.6$ subdiffusion sets in; this value is roughly similar for the spin transport at $\Delta = 1$. 
For larger anisotropy $\Delta>1$ (i.e. the strongly interacting regime, current scalings in Ref. \cite{Suppl}) we also find a diffusion-subdiffusion transition at a finite disorder strength, 
as shown in the inset of Fig. \ref{enerscalings}(b). 
However, this is entirely different to the case of spin transport, which becomes subdiffusive for much weaker (perhaps infinitesimal) disorder strengths in this regime of large anisotropy \cite{Znidaric:2016} . 
This effect is seen more conspicuously in Fig. \ref{Seebeck_currents_D175} for zero magnetization: for the same large anisotropy $\Delta>1$, energy diffusion is clearly discernable whereas spin transport is strongly subdiffusive. 
%
%That is, \textit{energy is transported diffusively whereas spin is transported subdiffusively in the strongly-interacting regime of the disordered spin chain}. 

Based on these NESS current exponents $\gamma$ extracted for a range of disorder and interaction strengths (see also additional data in Supplemental Information~\cite{Suppl}), we display a summary in the 
colour map in Fig.~\ref{enerdiagram_fig} on a two-dimensional landscape of $h$ vs. $\Delta$. The black line indicates the boundary between diffusive and subdiffusive transports for energy.
We already see a striking dissimilarity from that for spin transport: the phase boundary in the latter also linearly increases from the origin of this plot \textit{but} 
then buckles back towards the anisotropy-line around $\Delta\approx1$. Thus, spin transport for strongly anisotropic systems immediately becomes subdiffusive upon introducing disorder, while for energy transport the diffusive-subdiffusive boundary 
increases all the way at least up to a strong anisotropy of $\Delta = 4$ \cite{Suppl}. Consequently we can conclude that \textit{a large swathe of this landscape remains diffusive for energy transport in 
contrast to spin conduction.} 

%This last finding is in qualitative agreement with energy-quench studies in the $XXZ$ model \cite{Varma:2017JSTAT} where different $k$-modes were excited and allowed to relax; 
%however the exact phase boundary between diffusive and subdiffusive transport is not the same, quite likely to the much smaller system sizes 
%accessible in that study as well as the different nonequilibrium protocol considered here (i.e. the introduction of boundary baths).

\begin{figure}[t]
\centering
\includegraphics[width=0.42\textwidth]{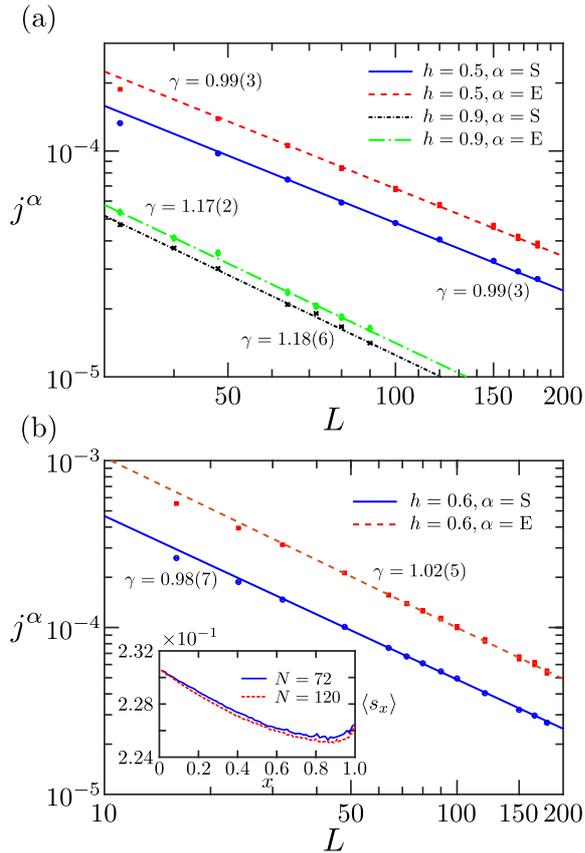}
\caption{NESS transport at finite magnetization. 
(a) Scaling of energy and spin currents for transport with $\Delta=1$ and two different disorder strengths, 
indicating diffusive ($h=0.5$) and subdiffusive ($h=0.9$) transport. 
%The results for $h=1.2$ are not plotted, but different fits resulted in $\nu=0.989(17),0.987(24)$ for spin and $\nu=1.007,1.018$ 
%for energy (close to $\nu=1.03$ found for zero magnetization).
%
(b) Scaling of energy and spin currents for transport with $\Delta=1.75$ and $h=0.6$, indicating diffusive transport. Inset: Magnetization profiles for different system sizes, with rescaled axis $x$.
}
\label{Seebeck_currents_D1}
\end{figure}

\textit{\underline{Transport at finite magnetization}}: In the zero magnetization sector spin and energy conduction were seen in Fig.~\ref{enerdiagram_fig}(b) to be different in nature for strong anisotropy $\Delta$; this is true also in the absence of disorder. However, in a clean system and in any finite magnetization sector, both spin and energy transports are ballistic for all values of $\Delta$, 
as known from conservation-law arguments~\cite{zotos:1997} and numerical calculations~\cite{louis:2003,ajisaka:2012,Mendoza:2015}. 
Whether both quantities show the same transport behavior in the presence of disorder is a natural question which we discuss in the following. 

For this scenario we impose a driving on the spin chain that induces a finite magnetization on every site, in addition to the temperature imbalance between the boundaries (see Ref.~\cite{Suppl} for details). 
This leads to a sizable spin current coexisting with the energy current, as shown in the main panels of Fig.~\ref{Seebeck_currents_D1}, and a corresponding magnetization gradient across the lattice, as depicted in the inset of Fig.~\ref{Seebeck_currents_D1}(b). The results of current scaling obtained for $\Delta=1$ in the presence of disorder are shown in Fig.~\ref{Seebeck_currents_D1}(a). 
As in the zero magnetization sector at the isotropic point, both spin and energy currents scale similarly as a function of disorder strength ($h=0.5$ and $h=0.9$ are displayed), 
signalling equal transport dynamics (diffusion or subdiffusion for the two $h$ values respectively) for energy and spin.

For larger anisotropies a similar conclusion holds, which is significantly different from the case of zero magnetization. 
Results for this strongly-interacting regime are shown in Fig.~\ref{Seebeck_currents_D1}(b), where for $h=0.6$ we have diffusive spin and energy transport, in stark contrast to the zero magnetization sector results of Fig. \ref{Seebeck_currents_D175}.
We note that for stronger disorder both spin and energy transport become simultaneously subdiffusive. The determination of the critical value at which this happens is left for a future study, but preliminary simulations indicate that for $h=1$ we have $\gamma > 1$, indicating subdiffusion.

The overarching conclusion of our work is therefore that the presence of conservation laws in the clean $XXZ$ limit seem to determine the dynamics of energy and spin in the weakly disordered limit. 
Weak disorder slows the transport by a ``single step'' from the clean limit: while the clean $XXZ$ always shows ballistic energy transport as a function of anisotropy, 
the spin transport varies substantially with anisotropy depending on the magnetization, thereby leading to differing dynamics of energy and spin even in the disordered regime as a function of anisotropy.

\textit{\underline{Experimental realization}}: Finally we comment on a very promising architecture for observing the discussed phenomena in the laboratory. 
In recent years a highly-controllable and optimally-measurable cold-atom scheme for experimentally studying nonequilibrium phenomena has been developed~\cite{Krinner:2017}, 
in which two unequal reservoirs of atoms are connected through a low-dimensional channel. With this setup particle currents induced by a chemical potential 
bias have been experimentally analyzed~\cite{Brantut:2012,Stadler:2012,Krinner:2015}, even in the presence of disorder generated by a speckle pattern projected onto the channel~\cite{Krinner:2013}. 
It has also been exploited to induce energy and thermoelectric transport by a temperature difference across the reservoirs~\cite{Brantut:2013}. 
More recently, by projecting optical barriers on top of the channel, a mesoscopic one-dimensional lattice was engineered, whose conduction properties were characterized in the presence of particle-particle 
interactions~\cite{Lebrat:2017}. By combining and extending these experimental techniques, it would be possible to induce particle and energy transport through 
disordered interacting lattices of tens of sites, enabling the predictions presented in our work to be verified. 

\textit{\underline{Conclusions}}: We have studied the nonequilibrium steady states of boundary driven interacting and disordered spin-1/2 chains. The drives were set up such that the boundary spins thermalized to density matrices given by different temperatures at 
each end, thereby setting up a temperature gradient, and an associated energy current. 
In the zero magnetization sector (and far from any localization physics) we found a regime of diffusive energy transport separated from a subdiffusive energy transport regime at finite values of disorder strength.
Moreover, for strong interactions there is a phase where energy is diffusive but the spin is subdiffusive.
However, at finite magnetization their NESS exponents (and corresponding dynamics) are found to be the same, whether at strong or weak anisotropies. 
Based on these we conclude that the conservation laws in the clean limit lead to the differing steady state exponents of energy and spin in the weakly disordered problem.

The authors would like to acknowledge the use of the University of Oxford Advanced Research Computing (ARC) facility in carrying out this work. http://dx.doi.org/10.5281/zenodo.22558. 
This research is partially funded by the European Research Council under the European Union's Seventh Framework Programme (FP7/2007-2013)/ERC Grant Agreement no. 319286 Q-MAC. 
This work was also supported by the EPSRC National Quantum Technology Hub in Networked Quantum Information Processing (NQIT) EP/M013243/1. 
JJM-A acknowledges financial support from Vice-Rectoría Investigaciones through UniAndes-2015 project \emph{Quantum control of nonequilibrium hybrid systems-Part II}. 
J.G. is supported by a SFI Royal Society University Research Fellowship.
A.S.\ is partially supported by a Google Faculty Award.
We acknowledge helpful discussions with A. Dymarsky, D. A. Huse, and V. Oganesyan.

\bibliography{disorder_bib}	

\clearpage

\renewcommand\thesection{S\arabic{section}}
\renewcommand\theequation{S\arabic{equation}}
\renewcommand\thefigure{S\arabic{figure}}
\setcounter{equation}{0}

\onecolumngrid
\begin{center}
{\Large Supplemental Material for \\ ``\Title''}
\end{center}

\twocolumngrid

\section{Simulation of two-site driving scheme} \label{driving_appendix}

In the following we describe how the two-site driving protocol for inducing a nonequilibrium NESS in a disordered lattice is implemented, and present details of its numerical simulation.

\subsection{Defining target states for energy and spin driving}

To study spin and energy transport across disordered spin lattices, we simulate the nonequilibrium configuration depicted in Fig.~\ref{setup_fig}(a). Here the two left-most (L) and right-most (R) sites are coupled to reservoirs of different temperature and chemical potential. Let $\beta_k$ and $\mu_k$ be the target inverse temperature and chemical potential characterizing the target state $\rho_k$ that reservoir $k$ tries to impose at the $k$ boundary ($k=\text{L,R}$). The corresponding grand-canonical state for $m>2$ sites is
\begin{align} \label{driving}
\begin{split}
\rho_k^{(m)}&=\exp\left(\beta_k\left(-H_k^{(m)}+\mu_k M^{(m)}\right)\right)/Z_k^{(m)},\\
Z_k^{(m)}&=\text{Tr}\left[\exp\left(\beta_k\left(-H_k^{(m)}+\mu_k M^{(m)}\right)\right)\right],
\end{split}
\end{align}
where $H_k^{(m)}$ is the Hamiltonian for the $m$ sites at boundary $k$, $M^{(m)}=\sum_{i=1}^ms_i^z$ is the magnetization operator for the $m$ sites, and $Z_k^{(m)}$ is the corresponding partition function. The two-site target state $\rho_k$ results from tracing out the $m-2$ most internal sites from $\rho_k^{(m)}$. Once the target states are defined for both left and right boundaries for a particular value of $m$ (we take $m=4$), the Lindblad dissipators $\mathcal{L}_k$ are built so that $\mathcal{L}_k(\rho_k)=0$, with the coupling strength between the chain and the reservoirs being $\Gamma=1$. Full details of how this is done are given in several references~\cite{Prosen:2009,Rossini:2010,Znidaric:2011c,Mendoza:2015}, which we summarize below. We emphasize that even though a microscopic derivation of master equation~\eqref{lindblad} with the described two-site driving might be quite challenging~\cite{breuer}, the transport properties we obtain in the bulk are independent of the details of such driving for the system sizes we take in our calculations~\cite{Znidaric:2016}. 

Several nonequilibrium configurations can be induced by choosing different values of $\beta_k$ and $\mu_k$. In our work we are interested in two cases. Firstly we consider energy transport for zero total magnetization, where $\beta_{\text{L}}<\beta_{\text{R}}$ and $\mu_{\text{L}}=\mu_{\text{R}}=0$. In particular, our simulations were performed with $\beta_\text{L}=4\times10^{-3}$ and $\beta_\text{R}=4\times10^{-2}$. This leads to a NESS in which an energy flow is established from left to right, with negligible net magnetization flow across the lattice. Secondly we discuss the case of a temperature imbalance beyond half filling, where $\beta_\text{L}<\beta_{\text{R}}$ and {$\beta_\text{L}\mu_{\text{L}}=\beta_\text{R}\mu_{\text{R}}>0$. Specifically, for our simulations we took the same temperature imbalance and $\beta_k\mu_{k}=1$. Here a significant spin current emerges in addition to the energy flow, which features a different nature to the spin transport at zero magnetization and no temperature imbalance~\cite{Znidaric:2016}, as discussed in the main text.

\subsection{Building the driving superoperator}

Now we briefly describe how to create a two-site bath superoperator $\mathcal{L}$ inducing a target state $\rho$ as the NESS of a pair of isolated sites, such that
\begin{equation}
\mathcal{L}(\rho)=0.
\end{equation}
We need to define $\mathcal{L}$ so that $\rho$ is its only eigenvector with zero eigenvalue, and that all the other eigenvalues are equal to $-1$. This leads to the fastest possible convergence to $\rho$~\cite{Prosen:2009}. For this, we first diagonalize the target state, $\rho=V^{\dagger}dV$, and obtain the ``diagonal" superoperator $\mathcal{L}^{\text{diag}}$ for which the diagonal matrix $d$ is the only zero-eigenvalue eigenstate,
\begin{equation}
\mathcal{L}^{\text{diag}}(d)=0.
\end{equation}
The matrix elements of the ``diagonal" superoperator are determined depending on the particular definition of the two-site operator-space basis $\Omega_n$ ($n=0,\ldots,15$). For example, for $\Omega_n=(\sigma_{n_1}\otimes\sigma_{n_2})/4$, with $\sigma_{n_{\alpha}}$ ($\alpha=1,2$) the Pauli matrices plus the identity ($n_{\alpha}=0,\ldots,3$), the exact form of $\mathcal{L}^{\text{diag}}$ is given in Refs.~\cite{Prosen:2009,Mendoza:2015} for two different orders of the Pauli basis.

Once $\mathcal{L}^{\text{diag}}$ is built, the total superoperator $\mathcal{L}$ is obtained through a rotation in operator space,
\begin{equation}
\mathcal{L}=R^{\dagger}\mathcal{L}^{\text{diag}}R.
\end{equation}
The elements of the rotation matrix $R$ are given by
\begin{equation}
R_{i,j}=\frac{1}{4}\text{tr}(V^{\dagger}\Omega_iV\Omega_j),
\end{equation}
thus being defined through $V$.

\subsection{Details on numerical simulation method}

\begin{figure*}[tp!]
\centering
\includegraphics[width=0.9\textwidth]{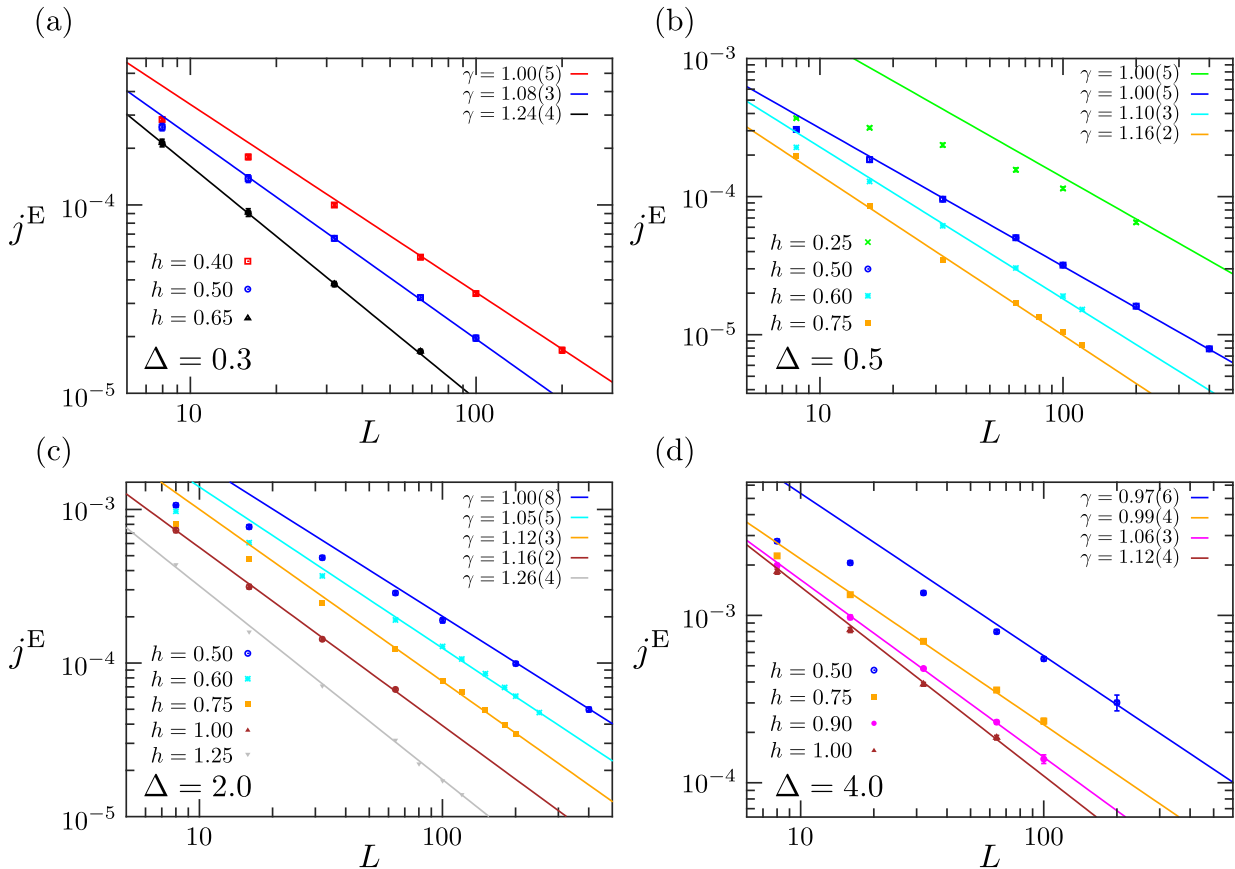}
\caption{Transport at zero magnetization, showing scaling of energy current in NESS for various anisotropies $\Delta$ and disorder strengths $h$. 
As in Fig. \ref{enerscalings}(b) of the main text, the disorder strength $h$ increases from top to bottom in each plot.  
%.
We find that stronger disorders are required for the onset of subdiffusion as we enter the Ising regime.
}
\label{enerscalings_supp}
\end{figure*}

To obtain the NESS of the system for each set of parameters, we consider $M$ realizations of the disordered magnetic field. For each realization $r$ we simulate the long-time evolution of the density matrix of the lattice, given by $\rho_{\infty}^{(r)}=\lim_{t\rightarrow\infty}\exp(\mathcal{L}^{(r)}t)\rho(0)$, with $\rho(0)$ the initial state. A unique NESS $\rho_{\infty}^{(r)}$ is obtained after evolving for a long-enough time from any $\rho(0)$, guaranteed due to the ergodicity of the bulk coherent dynamics~\cite{Prosen:2009}. We take $\rho(0)$ as a product state with homogeneous magnetization $\langle s_i^z\rangle=\beta_{k}\mu_{k}/4$, which is the value imposed by the driving on two isolated sites.

The time-evolution simulation is performed using the t-DMRG technique, which allows us to analyze spin chains of hundreds of lattice sites. The algorithm, implemented with the open-source Tensor Network Theory (TNT) library~\cite{tnt,tnt_review1}, is based on a Suzuki-Trotter decomposition~\cite{Suzuki:1990} of the Lindblad evolution operator. Here at any time $t$ the density matrix of the system $\rho^{(r)}(t)$ is described by a matrix product operator (MPO) with matrix dimension of up to $\chi=150$, and its time evolution is calculated by a sequence of two-site gates corresponding to applications of local evolution operators~\cite{Marko:2017ann}. This process is performed until the currents become homogeneous across the lattice, which indicates that the NESS has been reached. 

After the NESS is obtained for $M$ realizations (taking up to $M=200$ for small system sizes), the expectation values of interest are averaged over all of them, getting maximal statistical uncertainty of $\sigma(j^E)/\sqrt{M}\approx 2\%$ for the zero-magnetization sector and of $\approx 3-4\%$ for the harder-to-simulate nonzero magnetization transport, with $\sigma(j^E)$ the standard deviation of the NESS energy current when averaged over all the disorder realizations and across the lattice.  

\section{Additional current scaling}

In Fig. \ref{enerscalings}(b) of the main text, we showed that from the current scalings for different anisotropies $\Delta$ we may claim that (i) there is a transition from diffusive to subdiffusive 
transport of energy as a function of disorder strength; (ii) this critical disorder strength is far from the localization transition.
Similar conclusions also hold for other weaker and stronger anisotropies as shown in Fig. \ref{enerscalings_supp}.
However we note that as we increase the Ising anisotropy, the diffusive regime also increases i.e. a larger portion of the Hamiltonian-paramater space is diffusive before the onset of subdiffusion.
Thus we have a regime where spin is transported subdiffusively whereas the energy is transported diffusively; a particular instance is shown in Fig. \ref{Seebeck_currents_D175} where this phenomenon is seen 
more clearly.

Collecting all data from Fig. \ref{enerscalings} and Fig. \ref{enerscalings_supp} results in the dynamical phase diagram Fig. \ref{enerdiagram_fig}(b).

\begin{figure}[t]
\centering
\includegraphics[width=0.4\textwidth]{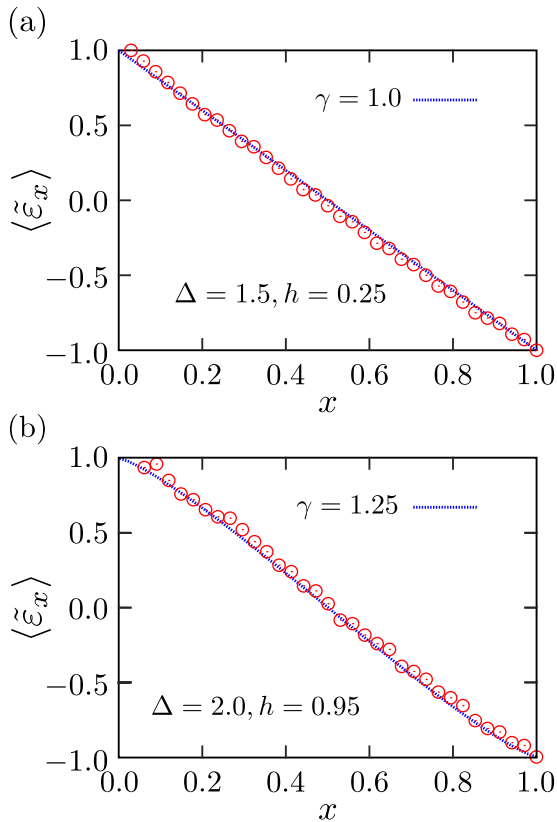}
%\caption{Energy profiles (averaged over disorder) of chains of several size with different types of NESS energy transport. 
%%
%(a) For $\Delta=1.5$ and $h=0.6$, showing collapse among various system sizes after rescaling the $x$-axis as $x=(j-1)/(L-2)$, for $j=1,\ldots,L-1$.
%%
%Lower panels: Energy profile ($L=64$) fits using spatially varying diffusion constant $D(x) \propto [x(1-x)]^{1-\gamma}$ for \textit{bulk} transport (20-30\% of either ends of the chain have been removed), 
%for various anisotropies and disorder strengths. (b) Naive NESS current scaling would give superdiffusive behaviour from finite-L which from the profile nicely shows diffusion instead. 
%(c) Here at large anisotropy and disorder the bulk still shows nice subdiffusive scaling despite 50\% of the edges being removed.}
\caption{Rescaled energy profiles averaged over disorder for $L=64$ and parameters leading to different type of energy transport (symbols), and fits using a spatially varying diffusion constant $D(x) \propto [x(1-x)]^{1-\gamma}$ for \textit{bulk} transport (dotted lines); 20-30\% of either ends of the chain has been removed. (a) Naive NESS current scaling would give superdiffusive behaviour from finite-L, which from the profile nicely shows diffusion instead. (b) Here at large anisotropy and disorder the bulk still shows nice subdiffusive scaling despite 50\% of the edges being removed. }
\label{enerprofiles_fig}
\end{figure} 

\section{Energy profiles and bulk transport}
A point to note from Fig.~\ref{enerscalings} and Fig.~\ref{enerscalings_supp} is that for very weak disorder fields $h=0.25$, the thermodynamic limit is reached only very slowly and fitting a straight line is 
problematic. 
Nevertheless we have drawn a $1/L$ diffusive fit that looks like it will asymptotically be parallel to the data. To better characterize the transport in this and similar cases, we fitted energy profiles to generalized diffusion equations with spatially 
dependent diffusion constants~\cite{Znidaric:2016}, as shown in Fig.~\ref{enerprofiles_fig}, where the slowest asymptoting (most nondiffusive i.e. naively superdiffusive fit) case of $h=0.25$, $\Delta=1.5$ clearly
shows diffusive scaling in the bulk (Fig. \ref{enerprofiles_fig}(a)). A $\gamma < 1$ fit to the energy profile cannot be reasonably made, thereby discounting superdiffusive transport for bulk physics. 
For $h=0.95$, $\Delta=4$ (Fig. \ref{enerprofiles_fig}(b)) where the data already converged (Fig. \ref{enerscalings_supp}) to subdiffusive transport, 
the energy profile provides a quantitatively consistent result (albeit a slightly different numerical value for the exponent).

\end{document}